\documentclass[jkps,twocolumn,fleqn,showpacs,showkeys]{revtex4}
\usepackage{graphicx}
\usepackage[dvipdfmx]{color}  
\usepackage{amssymb}
\usepackage{amsmath}
\usepackage{bm}

\newcommand{\ui}{{\rm i}} 
\newcommand{\bmr}{{\bm r}}

\newcommand{\bmq}{{\bm q}}

\newcommand{\bmk}{{\bm k}}
\newcommand{\bmK}{{\bm K}}
\newcommand{\bmp}{{\bm p}}
\newcommand{\kB}{k_{\rm B}}

\begin{document}
\setcounter{page}{0}
\title[]{Linear-response theory of the longitudinal spin Seebeck effect} 
\author{Hiroto \surname{Adachi}}
\email{adachi.hiroto@jaea.go.jp}
\author{Sadamichi \surname{Maekawa}} 
\affiliation{Advanced Science Research Center, Japan Atomic Energy Agency, 
Tokai 319-1195, Japan}


\begin{abstract} 
We theoretically investigate the {\it longitudinal} spin Seebeck effect, in which the spin current is injected from a ferromagnet into an attached nonmagnetic metal in a direction {\it parallel} to the temperature gradient. Using the fact that the phonon heat current flows intensely into the attached nonmagnetic metal in this particular configuration, we show that the sign of the spin injection signal in the longitudinal spin Seebeck effect can be opposite to that in the conventional transverse spin Seebeck effect when the electron-phonon interaction in the nonmagnetic metal is sufficiently large. Our linear-response approach can explain the sign reversal of the spin injection signal recently observed in the longitudinal spin Seebeck effect. 

\end{abstract}



\maketitle

\section{INTRODUCTION}

Because of the desire to deal with heating problems in modern spintronic devices, there has been an increasing interest in investigating thermal effects in spintronics. A new subfield ``spin caloritronics''~\cite{Bauer12} aims to understand the basic physics behind the interplay of spin and heat.  One of the central issues in spin caloritronics is the newly discovered thermo-spin phenomenon termed spin Seebeck effect~\cite{Uchida08}, which enables the thermal injection of spin currents from a ferromagnet into attached nonmagnetic metals over a macroscopic scale of several millimeters. The spin Seebeck effect is now established as a universal aspect of ferromagnets because this phenomenon is observed in various materials ranging from the metallic ferromagnets Ni$_{81}$Fe$_{19}$~\cite{Uchida08} and Co$_2$MnSi~\cite{Bosu11},  to the semiconducting ferromagnet (Ga,Mn)As~\cite{Jaworski10}, to the insulating magnets LaY$_2$Fe$_5$O$_{12}$~\cite{Uchida10a}. 

\begin{figure}[t]
\begin{center}
\includegraphics[width=6cm]{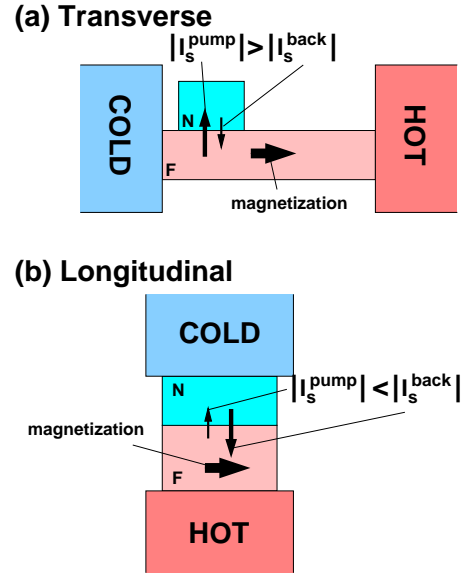} 
\end{center}
\caption{(Color online) Schematic view of the experimental setup for (a) the transverse spin Seebeck effect and (b) the longitudinal spin Seebeck effect. 
}\label{fig1_ICM2012Adachi}
\end{figure}

It is important to note that the above experiments~\cite{Uchida08,Bosu11,Jaworski10,Uchida10a} were performed in a configuration of the {\it transverse} spin Seebeck effect, in which the direction of the thermal spin injection into the attached nonmagnetic metal is {\it perpendicular} to the temperature gradient [Fig.~\ref{fig1_ICM2012Adachi} (a)]. Recently, another type of spin Seebeck effect called the {\it longitudinal} spin Seebeck effect~\cite{Uchida10b,Uchida10c} is reported, in which the direction of the thermal spin injection into the nonmagnetic metal is {\it parallel} to the temperature gradient [Fig.~\ref{fig1_ICM2012Adachi} (b)]. Whereas the longitudinal spin Seebeck effect is well defined only for the use of an {\it insulating} ferromagnet due to the parasitic contribution from the anomalous Nernst effect~\cite{Huang11,Weiler12}, it has several attractive features: (i) it is substrate free, (ii) the configuration is much simpler than that of the transverse spin Seebeck effect, and (iii) it can be of wide application because it allows the use of bulk samples. 

Another pronounced feature of the longitudinal spin Seebeck effect is that the sign of the spin injection signal is opposite to that in the transverse spin Seebeck effect~\cite{Uchida10b,Uchida10c}. Physically, the longitudinal spin Seebeck effect is distinguished from the transverse spin Seebeck effect by the fact that the attached nonmagnetic metal is in contact with the heat bath in the longitudinal setup, while the attached nonmagnetic metal is out of contact with the heat bath in the transverse setup. This brings about a clear difference that the heat current intensely flows into the attached nonmagnetic metal in the case of the longitudinal spin Seebeck effect, whereas it does not in the case of the transverse spin Seebeck effect. It is obvious that theory of magnon-driven spin Seebeck effect~\cite{Xiao10} fails to explain the situation in question. 

In this paper, by employing linear-response theory of the spin Seebeck effect~\cite{Adachi11} and using the importance of the phonon-drag process in the spin Seebeck effect~\cite{Adachi10}, we show that the sign of the spin injection signal in the longitudinal spin Seebeck effect can be opposite to that in the conventional transverse spin Seebeck effect when the electron-phonon interaction in the attached nonmagnetic metal is sufficiently large. The key in our discussion is the aforementioned difference in the position of the attached nonmagnetic metal between the longitudinal setup and the transverse setup.

\section{Phenomenology of the longitudinal spin Seebeck effect \label{sec:phenomenology}} 

Let us begin with the phenomenology of the longitudinal spin Seebeck effect. In Fig.~\ref{fig1_ICM2012Adachi}, a hybrid structure of a ferromagnet ($F$) and a nonmagnetic metal ($N$) is placed under a temperature gradient. The central quantity that characterizes the spin Seebeck effect is the spin current $I_s$ injected into $N$. As explained in detail in Ref.~\cite{Adachi12}, the spin Seebeck effect is a thermal spin injection by localized spins, and the injected spin current has two contributions, 
\begin{equation}
  I_s = I_s^{\rm pump}- I_s^{\rm back}, 
\end{equation}
where $I_s^{\rm pump}$ (the so-called pumping component) represents the spin current pumped into $N$ by the thermal fluctuations of localized spins in $F$ , while $I_s^{\rm back}$ (the so-called backflow component) represents the spin current coming back into $F$ by the thermal fluctuations of the spin accumulations in $N$. We now focus on the spin current injected into $N$ which is located close to the {\it cold} reservoir.

In the case of the conventional transverse spin Seebeck effect, the magnitude of the pumping component $I_s^{\rm pump}$ is greater than that of the backflow component $I_s^{\rm pump}$ [Fig.~\ref{fig1_ICM2012Adachi}(a)]. In contrast, the magnitude of $I_s^{\rm pump}$ is less than that of the backflow component $I_s^{\rm pump}$ in the case of the longitudinal spin Seebeck effect [Fig.~\ref{fig1_ICM2012Adachi}(b)]. Note that, because magnons carry minus spin 1, both the pumping and backflow components have a negative sign.

This difference can be explained phenomenologically on the basis of the following conditions: (i) most of the heat current in the $F$/$N$ hybrid system at room temperature is carried by phonons (see Ref.~\cite{Slack71} in the case of yttrium iron garnet), and (ii) the interaction between the phonons and the spin accumulation in $N$ is much stronger than the 
magnon-phonon interaction in $F$. 

First, recall that the pumping and backflow components can be expressed 
as follows~\cite{Adachi12}: 
\begin{eqnarray}
I_s^{\rm pump} &=& -G_s k_B T_F^*, \\
I_s^{\rm back} &=& -G_s k_B T_N^*, 
\end{eqnarray}
where $T_F^*$ and $T_N^*$ are the effective temperature of the magnon in $F$ and the spin accumulation in $N$. Here, $G_{s}= J^2_{\rm sd} \chi_{N} \tau_{\rm sf}/\hbar$ with $J_{\rm sd}$, $\chi_N$, and $\tau_{\rm sf}$ being the $s$-$d$ interaction at the interface, the paramagnetic susceptibility in $N$, and the spin-flip relaxation time in $N$, respectively. The negative sign before $G_s$ arises from the fact that the magnon carries spin $-1$. In the longitudinal spin Seebeck experiment, the nonmagnetic metal $N$ is in direct contact with the heat bath, and thereby is exposed to the flow of the phonon heat current due to condition (i). Then, because of condition (ii), spin accumulation in $N$ is heated up faster than the magnons in the ferromagnet $F$, and the resultant effective temperature of the spin accumulation in $N$ increases above that of the magnons in $F$. In the conventional transverse spin Seebeck setup, by contrast, the nonmagnetic metal $N$ is out of contact with the heat bath and the phonon heat current does not flow through the nonmagnetic metal $N$, while the ferromagnet $F$ is in contact with the heat bath, resulting in an increase in the effective magnon temperature in $F$. Therefore, in this case, the effective temperature of the spin accumulation in $N$ is lower than that of the magnons in $F$. This difference can explain the sign reversal of the spin Seebeck effect signal between the longitudinal setup and the conventional transverse setup.

\section{Linear-response Formulation} 
\begin{figure}[t]
\begin{center}
\includegraphics[width=8.0cm]{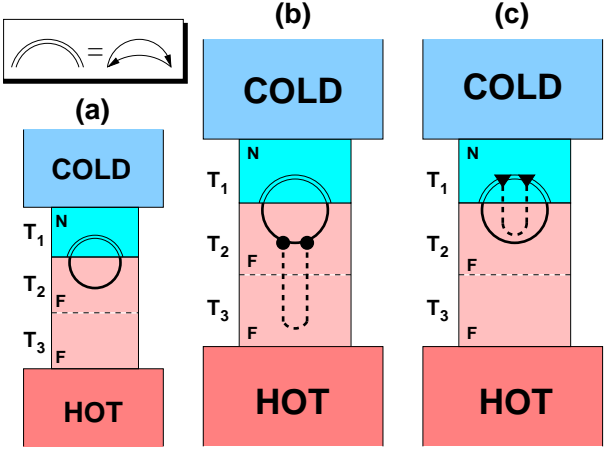}
\end{center}

\caption{(Color online) Diagrammatic representation of 
the spin Seebeck effect in the longitudinal configuration. 
The double solid line (the dashed line) represents 
the propagator of the itinerant spin density (the phonon). 
The solid circle (solid triangle) denotes the interaction vertex 
$\Gamma_{{\bmK},\bmq}$ ($\Upsilon_{{\bmK},\bmk}$) between 
the magnon and the phonon (the itinerant spin density and the phonon). 
Here, $T_1 < T_2 < T_3$. 
}\label{fig2_ICM2012Adachi}
\end{figure}

In this section we review the linear-response formalism of the 
spin Seebeck effect developed in Ref.~\cite{Adachi11}. 
In the next section, this formalism is employed to evaluate the 
longitudinal spin Seebeck effect. 
We use a model shown in Fig.~\ref{fig2_ICM2012Adachi}, in which 
the localized spins in $F$ are interacting with the spin accumulation in $N$ 
through the $s$-$d$ exchange interaction $J_{\rm sd}$ at the interface. 
In our approach, the spin accumulation is modeled as a 
nonequilibrium itinerant spin density ${\bm s}$. 

As in Ref.~\cite{Adachi11}, the spin current $I_s$ injected into the nonmagnetic metal $N$ is calculated as 
$I_s(t) = -\sum_{\bmq,\bmk}  \frac{4{\cal J}^{\bmk+\bmq}_{\rm sd} \sqrt{S_0} } 
{\sqrt{2 N_F N_N} \hbar} 
{\rm Re} C^{<}_{\bmk,\bmq}(t,t)$, 
where $N_F$ ($N_N$) is the number of lattice sites 
in $F$ ($N$), $S_0$ is the size of the localized spins in $F$, 
and ${\cal J}_{\rm sd}^{\bmk+\bmq}$ is the Fourier transform of the 
$s$-$d$ interaction at the $F/N$ interface. 
Here, 
$C^{<}_{\bmk,\bmq}(t,t') = - \ui \langle  a^+_\bmq(t') s^-_\bmk(t) \rangle $ 
measures the correlation between 
the magnon operator $a_\bmq^+$ in $F$ and the itinerant spin-density operator 
$s^-_\bmk= (s^x_\bmk- \ui s^y_\bmk)/2$ in $N$. 
Note that the time dependence of $I_s(t)$ vanishes in the steady state 
and it is hereafter discarded. 
Introducing the frequency representation 
$C^{<}_{\bmk,\bmq}(t-t') = 
\int_{-\infty}^\infty \frac{d \omega}{2 \pi} 
{C}^{<}_{\bmq,\bmk}(\omega) e^{- \ui \omega (t-t')}$ 
and adopting the representation~\cite{Larkin75} 
$\check{C} 
= \left({{C^{R}, C^{K}} \atop {0 \;\;\;  ,C^{A}}} \right)$
as well as using the relation $C^{<}= \frac{1}{2} [C^{K}- C^{R} + C^{A}]$, 
we obtain 
\begin{eqnarray}
  I_s &=&   \sum_{\bmq,\bmk} 
   \frac{-2{\cal J}^{\bmk-\bmq}_{\rm sd} \sqrt{S_0} }
  {\sqrt{2 N_F N_N} \hbar} 
  \int_{-\infty}^\infty \frac{d \omega}{2 \pi} 
  {\rm Re} C^{K}_{\bmk,\bmq}(\omega) 
  \label{Eq:I_s01}
\end{eqnarray}
for the spin current $I_s$ in the steady state~\cite{com1}.

Up to the lowest order in the $s$-$d$ interaction $J_{\rm sd}$, the interface correlation function $\check{C}$ appearing in Eq.~(\ref{Eq:I_s01}) is generally expressed as 
\begin{eqnarray}
  \check{C}_{\bmk,\bmq} (\omega) &=& 
  \frac{{\cal J}^{\bmk-\bmq}_{\rm sd} \sqrt{S_0} }{\sqrt{N_N N_F} \hbar}
  \check{\underline{\chi}}_\bmk(\omega) \check{\underline{X}}_\bmq(\omega), 
  \label{Eq:C-func01}
\end{eqnarray}
where $\check{\underline{X}}_\bmq(\omega)= \check{X}_\bmq(\omega) 
+ \delta \check{X}_\bmq(\omega)$ 
is the renormalized magnon propagator with the bare component 
$\check{X}_\bmq(\omega)$, 
and $\check{\underline{\chi}}_{\bmk}(\omega) = \check{\chi}_{\bmk}(\omega) 
+ \delta \check{\chi}_{\bmk}(\omega)$ 
is the renormalized spin-density propagator with the bare component 
$\check{\chi}_{\bmk}(\omega)$. 
The bare magnon propagator satisfies the equilibrium condition: 
\begin{eqnarray}
X^A_\bmq(\omega) = [X^R_\bmq(\omega)]^*, && 
X^K_\bmq(\omega) = 2 \ui \, {\rm Im} X^R_\bmq(\omega) 
\coth(\tfrac{\hbar \omega}{2 \kB T}), \nonumber \\
\label{Eq:X-eq01} 
\end{eqnarray} 
where the retarded component is given by 
$X^R_\bmq(\omega)= (\omega-\widetilde{\omega}_\bmq+ \ui \alpha \omega)^{-1}$ 
with $\widetilde{\omega}_\bmq = \gamma H_0 + \omega_\bmq$ 
being the magnon frequency for uniform mode $\gamma H_0$ 
and exchange mode $\omega_\bmq = D_{\rm ex}q^2/\hbar$. 
Likewise, the bare spin-density propagator 
satisfies the local equilibrium condition: 
\begin{eqnarray}
\chi^A_\bmk(\omega)= [\chi^R_\bmk(\omega)]^*,&&
\chi^K_\bmk(\omega)= 2 \ui \, {\rm Im} \chi^R_\bmk(\omega)
\coth(\tfrac{\hbar \omega}{2 \kB T} ), \nonumber \\
\label{Eq:chi-eq01} 
\end{eqnarray} 
where the retarded component of $\check{\chi}_{\bmk}(\omega)$ is given by 
$\chi^R_\bmk(\omega)
= \chi_N/(1+ \lambda_{\rm sf}^2 k^2 - \ui \omega \tau_{\rm sf})$ 
with $\lambda_{\rm sf}$ being the spin diffusion length. 

\section{Calculation of the longitudinal spin Seebeck effect} 

In this section we present a linear-response calculation of the 
longitudinal spin Seebeck effect and justify the phenomenological picture 
presented in Sec.~\ref{sec:phenomenology}. 
The spin current $I_s$ injected into $N$ due to the longitudinal spin Seebeck effect 
is composed of three terms, 
\begin{equation}
I_s = I_s(a)+ I_s(b)+ I_s(c), 
\label{Eq:Is-sum}
\end{equation}
where $I_s(a)$, $I_s(b)$, and $I_s(c)$ correspond to 
diagram (a), (b), and (c) in Fig.~\ref{fig2_ICM2012Adachi}. 
Below we show that each term has the following sign: 
\begin{equation}
  I_s(a)< 0, \;\; I_s(b)< 0, \;\;  I_s(c) > 0. 
  \label{Eq:Is-sign}
\end{equation}

First, let us consider diagram (a). 
This can be calculated by setting 
$\check{\underline{X}}_\bmq(\omega) \to \check{X}_\bmq(\omega)$ 
and $\check{\underline{\chi}}_{\bmk}(\omega) \to \check{\chi}_{\bmk}(\omega)$ 
in Eq.~(\ref{Eq:C-func01}), 
which was already done in Ref.~\cite{Adachi11} with the injected 
spin current given by (see Eq.~(12) therein) 
\begin{eqnarray}
  I_s(a) &=& 
  \frac{4 N_{\rm int} {J}_{\rm sd}^2S_0^2 }{\sqrt{2}\hbar^2 N_{N} N_{F}} 
   \sum_{\bmq,\bmk} 
   \int_{\omega} 
    {\rm Im} \chi_{\bmk}^{R}(\omega) 
    {\rm Im} X_{\bmq}^{R}(\omega)  \nonumber \\ 
    && \quad \times 
    \left[ \coth(\tfrac{\hbar \omega}{2 \kB T_{2}}) 
         - \coth(\tfrac{\hbar \omega}{2 \kB T_{1}})   \right], 
      \label{Eq:Is-a01} 
\end{eqnarray}
where we have introduced the shorthand notation 
$\int_\omega = \int_{-\infty}^{\infty} \frac{d \omega}{2 \pi}$, 
and $N_{\rm int}$ is the number of localized spins at the $N$/$F$ interface. 
Note that $I_s(a)$ has a negative value due to $T_2 > T_1$. 

Next, let us consider diagram (b). 
In this process, the localized spins in $F$ is excited by the nonequilibrium 
phonon driven by the temperature gradient in $F$, hence this corresponds to 
the phonon-drag process~\cite{Adachi10}. 
Evaluation of the diagram was already given in Ref.~\cite{Adachi10}, 
although the calculation 
is lengthy and tedious (see the supplemental material therein). 
In short, this term can be calculated by setting in Eq.~(\ref{Eq:C-func01}) 
$\check{\underline{X}}_\bmq(\omega) \to 
\delta \check{X}^{n \mathchar`-eq}_\bmq(\omega)=
({0, \atop 0,} 
{{\delta X_\bmq^{n \mathchar`-eq,K} } \atop 0}) $ 
with its Keldysh component given by 
\begin{eqnarray} 
  \delta {X}^{n \mathchar`-eq,K}_\bmq (\omega) 
  &=&
  -2 \sum_{\bmK} \frac{\Gamma_{{\bmK},\bmq}^2}{2N_F} 
  \int_\nu \delta D^{n \mathchar`-eq,K}_\bmK(\nu)   
  {\rm Im}X^R_{\bmq_-}(\omega_-)  \nonumber \\
  &\times&  
  |X^R_\bmq(\omega)|^2  
  \left[ \coth(\tfrac{\hbar \omega_-}{2 \kB T_{2}}) ]
  - \coth(\tfrac{\hbar \omega}{2 \kB T_{2}})   \right], \nonumber \\
    \label{Eq:dX_noneq02}
\end{eqnarray}
where we have introduced shorthand notations  
$\omega_-= \omega-\nu$, $\bmq_-= \bmq- \bmK$, and 
$\int_\nu = \int_{-\infty}^\infty \frac{d \nu}{2 \pi}$, 
and 
\begin{equation}
\Gamma_{{\bmK},\bmq} = 
{g}_{\rm m-p} 
\sqrt{\frac{ \hbar \nu_\bmK}{2M_{\rm ion}v_{\rm ph}^2}} 
\label{Eq:Gamma} 
\end{equation}
is the magnon-phonon interaction vertex. 
Here, 
$\nu_\bmK$, $v_{\rm ph}$ and $M_{\rm ion}$ are the phonon frequency, 
phonon velocity and the ion mass, respectively, 
and the strength of the magnon-phonon coupling is given by 
$g_{\rm m-p}=|a_S {\bm \nabla} J_{\rm ex}| ({\omega_\bmq}/{J_{\rm ex}}) $ with the 
exchange interaction $J_{\rm ex}$. 
In Eq.~(\ref{Eq:dX_noneq02}), 
\begin{eqnarray}
  \delta {D}_\bmK^{n \mathchar`-eq,K} (\nu) &=& 
  2 \ui \sum_{\bmK'}  \frac{|\Omega_{\rm ph}^{\bmK+\bmK'}|^2}{N_F^2} 
  {\rm Im}{D}^R_{\bmK'}(\nu) \nonumber   |{D}^R_\bmK(\nu)|^2  \\
  &\times &
  \big[ \coth(\tfrac{\hbar \nu}{2 \kB T_{3}}) 
  - \coth(\tfrac{\hbar \nu}{2 \kB T_{2}}) \big] 
  \label{Eq:dD_noneq01}
\end{eqnarray}
is the Keldysh component of the 
nonequilibrium phonon propagator. 
Here ${\Omega}_{\rm ph}^{\bmK+\bmK'}$ is the Fourier transform of 
${\Omega}_{\rm ph}(\bmr)= \Omega_0 
\sum_{\bmr_0 \in F/F \mathchar`-{\rm interface}} 
a_S^3 \delta(\bmr-\bmr_0)$ 
with $\Omega_0=\sqrt{2K_{\rm ph}/M_{\rm ion}}$, $K_{\rm ph}$ is the elastic constant, 
and $a_S^3$ is the cell volume of the ferromagnet. 
Putting these expressions into Eq.~(\ref{Eq:C-func01}) and after some algebra, 
we finally obtain 
\begin{eqnarray}
  I_{s}(b) &=& 
  \frac{-L}{N_N N_F^3} \sum_{\bmk,\bmq,\bmK,\bmK'}
  (\Gamma_{{\bmK},\bmq})^2 
  \int_\nu A_{\bmk,\bmq}(\nu) |{D}^R_\bmK(\nu)|^2 \nonumber \\
      & & \times 
      {\rm Im}{D}^R_{\bmK'}(\nu) 
      \big[ \coth(\tfrac{\hbar \nu}{2 \kB T_{3}}) 
        - \coth(\tfrac{\hbar \nu}{2 \kB T_{2}}) \big] ,    
      \label{Eq:Is-b01}
\end{eqnarray}
where 
$D^R_\bmK(\nu)= 
(\nu-\nu_\bmK + \ui /\tau_{\rm ph})^{-1}- 
(\nu+\nu_\bmK+ \ui/\tau_{\rm ph})^{-1}$ 
is the retarded component of the phonon propagator with the 
phonon lifetime $\tau_{\rm ph}$, 
$L= \sqrt{2} (J^2_{sd}S_0) \Omega_0^2 N_{\rm int} N'_{\rm int}/N_F$ 
with $N'_{\rm int}$ being 
the number of lattice sites at the $F$/$F$ interface, 
and $A_{\bmk,\bmq}(\nu)$ is defined by 
\begin{eqnarray}
  A_{\bmk,\bmq}(\nu) &=& 
  \int_\omega {\rm Im}\chi^R_\bmk (\omega) 
      {\rm Im}X^R_{\bmq_-}(\omega_-) \hspace{3cm}\nonumber \\
  &\times& |X^R_\bmq(\omega)|^2 
      [\coth(\tfrac{\hbar \omega_-}{2 \kB T_{2}}) 
        - \coth(\tfrac{\hbar \omega}{2 \kB T_{2}})]. 
\end{eqnarray}
Note that only the even component 
of $A_{\bmk,\bmq}(\nu)$ as a function of $\nu$ 
gives a non-vanishing contribution 
to Eq.~(\ref{Eq:Is-b01}). 
Because the even component of $A_{\bmk,\bmq}(\nu)$ is negative definite 
as well as 
${\rm Im}{D}_\bmK^R(\nu) 
[ \coth(\tfrac{\hbar \nu}{2 \kB T_{3}}) 
- \coth(\tfrac{\hbar \nu}{2 \kB T_{2}}) ]$ 
in Eq.~(\ref{Eq:Is-b01}), 
$I_s(b)$ has a negative value. 

Finally, let us consider diagram (c). 
Repeating essentially the same procedure in evaluating diagram (b), 
we obtain 
\begin{eqnarray}
  I_{s}(c) &=& 
  \frac{L'}{N_N N_F^3} \sum_{\bmk,\bmq,\bmK,\bmK'}
  (\Upsilon_{{\bmK},\bmk})^2 
  \int_\nu B_{\bmk,\bmq}(\nu) |\widetilde{D}^R_\bmK(\nu)|^2 \nonumber \\
      & & \times 
      {\rm Im}{D}^R_{\bmK'}(\nu) 
      \big[ \coth(\tfrac{\hbar \nu}{2 \kB T_{2}}) 
        - \coth(\tfrac{\hbar \nu}{2 \kB T_{1}}) \big] ,    
      \label{Eq:Is-c01}
\end{eqnarray}
where $L'= \sqrt{2} (J^2_{sd}S_0) \Omega_0^2 N^2_{\rm int} /N_F$, 
$\widetilde{D}^R_\bmK(\nu)$ denotes the phonon propagator 
in $N$.
In the above equation, the coupling between 
the itinerant spin density and the phonon in $N$ 
is given by 
\begin{equation}
\Upsilon_{\bmK,\bmk} 
\approx 
g_{\rm s-p} 
\sqrt{\frac{ \hbar \nu_\bmK}{2M_{\rm ion}v_{\rm ph}^2}}, 
\label{Eq:Upsilon}
\end{equation}
where $g_{\rm s-p} \approx 
|a {\bm \nabla} t_{\rm hop}| U^2N'(0)$ 
with $a$ and $t_{\rm hop}$ being the lattice spacing and the hopping integral 
of the nonmagnetic metal $N$. 
In Eq.~(\ref{Eq:Is-c01}), $B_{\bmk,\bmq}(\nu)$ is defined by 
\begin{eqnarray}
  B_{\bmk,\bmq}(\nu) &=& 
  \int_\omega {\rm Im}\chi^R_\bmk (\omega) {\rm Im}X^R_{\bmq_-}(\omega_-) 
  |\chi^R_\bmq(\omega)|^2 \nonumber \\
  &&\times 
      [\coth(\tfrac{\hbar (\omega_-}{2 \kB T_{1}}) 
        - \coth(\tfrac{\hbar \omega}{2 \kB T_{1}})]. 
\end{eqnarray}
Note that as in Eq.~(\ref{Eq:Is-b01}), only the even-in-$\nu$ component of 
$B_{\bmk,\bmq}(\nu)$ gives non-vanishing contribution to Eq.~(\ref{Eq:Is-c01}). 
Then, because the even-in-$\nu$ component of $B_{\bmk,\bmq}(\nu)$ is 
negative definite, $I_s(c)$ has a positive value.

\section{discussion}

In the previous section, we proved 
that $I_s(a)$ and $I_s(b)$ have the same sign, whereas $I_s(c)$ have 
the opposite sign [Eq.~(\ref{Eq:Is-sign})]. 
Then, if $I_s(c)$ is dominant in Eq.~(\ref{Eq:Is-sum}), 
it means that the sign of the longitudinal spin Seebeck effect 
can be opposite to that of the transverse spin Seebeck effect, 
since the sign of $I_s(a)$ and $I_s(b)$ is the same as that of the 
transverse spin Seebeck effect (see Refs.~\cite{Adachi11} and~\cite{Adachi10}). 
Because $I_s(a)$ and $I_s(b)$ are considered to have the same 
magnitude at room temperature (see Fig.~3 in Ref.~\cite{Adachi10}), 
we here compare the magnitude of $I_s(b)$ and $I_s(c)$. 

The key quantities determining the magnitude of $I_s(b)$ and $I_s(c)$ are 
the interaction vertex $\Gamma_{{\bmK},\bmq}$ between magnons and phonons 
[solid circle in Fig.~\ref{fig2_ICM2012Adachi} and Eq.~(\ref{Eq:Gamma})] and 
the interaction vertex $\Upsilon_{{\bmK},\bmk}$ between spin 
accumulation and phonons [solid triangle in Fig.~\ref{fig2_ICM2012Adachi} 
and Eq.~(\ref{Eq:Upsilon})]. 
The magnitude of these couplings is roughly given by 
$g_{\rm m-p} \approx |a_S {\bm \nabla} J_{\rm ex}| (\omega_\bmq/J_{\rm ex})$ and 
$g_{\rm s-p} \approx 
|a {\bm \nabla} t_{\rm hop}| U^2N'(0)$, 
where $a$, $t_{\rm hop}$, $U$, $N'(0)$ are the lattice spacing, the hopping integral, 
the strength of the Coulomb repulsion, 
and the strength of the particle-hole asymmetry in the nonmagnetic metal $N$. 
For materials with a relatively large Coulomb repulsion $U$~\cite{Gunnarsson76} 
and moderate strength of particle-hole asymmetry $N'(0)$~\cite{Moore73} such as Pt, 
we expect a situation 
$g_{\rm s-p} > g_{\rm m-p}$, 
which then explains the sign reversal 
of the spin injection signal in the longitudinal spin Seebeck effect 
due to Eqs.~(\ref{Eq:Is-sum}) and (\ref{Eq:Is-sign}). 
From these considerations, we conclude 
that this happens in the longitudinal spin Seebeck effect 
reported in Refs.~\cite{Uchida10b} and~\cite{Uchida10c}. 

\section{Conclusion}
In this paper we have developed linear-response theory of the longitudinal 
spin Seebeck effect. 
We have shown that 
the sign of the spin injection signal in the longitudinal 
spin Seebeck effect can be opposite to that in the conventional 
transverse spin Seebeck effect 
when the interaction between the spin accumulation and the phonon 
in the attached nonmagnetic metal is sufficiently stronger than 
the interaction between the magnon and the phonon in the ferromagnet. 
The linear-response approach presented in this paper 
can explain the sign reversal of the 
spin injection signal recently observed in the longitudinal 
spin Seebeck effect~\cite{Uchida10b,Uchida10c}.

\begin{figure}[t]
\begin{center}
\includegraphics[width=5.0cm]{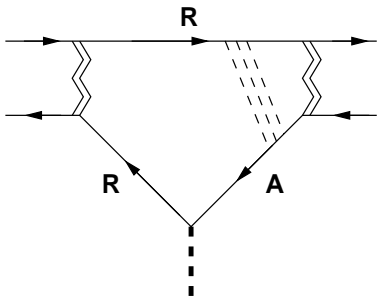}
\end{center}
\caption{Diagram corresponding to the interaction vertex $\Upsilon_{{\bmK},\bmk}$ 
between the itinerant spin density and the phonon. The solid line, 
the wavy line, the dashed line and the triple dashed line 
represent the electron propagator, the Coulomb repulsion, the phonon propagator, 
and the diffuson propagator, respectively. 
``R'' or ``A'' means the retarded or advanced component of the 
electron Green's function. 
}\label{fig3_ICM2012Adachi}
\end{figure}

\begin{acknowledgments}
The authors would like to thank K. Uchida and E. Saitoh for helpful 
discussions, and gratefully acknowledge support by 
a Grant-in-Aid for Scientific Research from MEXT, Japan. 
\end{acknowledgments}

\appendix

\section{Calculation of the vertex $\Upsilon$\label{Sec:appendix}}
In this Appendix, we evaluate the interaction vertex $\Upsilon_{\bmK,\bmk}$ 
(solid triangle in Fig.~\ref{fig2_ICM2012Adachi}) 
between the itinerant spin density and the phonon in the nonmagnetic metal $N$. 
We assume that the nonmagnetic metal $N$ has a moderately large Stoner 
enhancement factor, and for conduction electrons in $N$ we use a model 
described by Ref.~\cite{Kawabata74}, and assume an elastic impurity scattering 
as well. 
The interaction vertex before integrating out the fermionic degrees of freedom 
is shown in Fig.~\ref{fig3_ICM2012Adachi}. 
The building block of this diagram is given by a triangle 
\begin{eqnarray}
  {\cal T} &=& 
  \int_\epsilon 
  \left[ \tanh(\tfrac{\hbar (\epsilon- \omega)}{2 \kB T}) 
    - \tanh(\tfrac{\hbar \epsilon}{2 \kB T})   \right]  \nonumber \\
  && \hspace{1cm}\times  \int_{\bmp} {\rm Im}G^R_{\bmp-\bmk}(\epsilon -\omega ) 
  G^R_{\bm p}(\epsilon) G^A_{\bm p}(\epsilon), 
\end{eqnarray}
where 
$G^{R/A}_{\bm p}(\epsilon) = (\epsilon- \epsilon_\bmp \pm \ui /\tau)^{-1}$ 
is the electron Green's function with the electron's lifetime $\tau$. 
This diagram can be evaluated to be 
${\cal T} \approx N'(0) \omega \tau $ 
as was done in Ref.~\cite{Kamenev95}. 
After the inclusion of a diffuson vertex correction 
(triple dashed ladder in Fig.~\ref{fig3_ICM2012Adachi}) which is important 
in a realistic diffusive situation, 
we obtain 
\begin{equation}
  {\cal T} \approx N'(0) \frac{(\omega \tau)^2}{(Dk^2 \tau )^2+ (\omega \tau)^2}, 
\end{equation}
where $D$ is the diffusion constant. 
The dominant contribution comes from the dynamical region 
$\omega \gg D k^2$ and in this case we approximately have 
${\cal T} \approx N'(0)$. 
By attaching two Coulomb repulsion $U$ 
and one electron-phonon interaction 
$\approx |a {\bm \nabla} t_{\rm hop} |
\sqrt{\frac{ \hbar \nu_\bmK}{2M_{\rm ion}v_{\rm ph}^2}} $~\cite{Walker01} 
coming from each vertex, we finally obtain Eq.~(\ref{Eq:Upsilon}).


\begin{references}
\bibitem{com1}
  Definition of the magnon propagator in this paper differs 
  from that in Ref.~\cite{Adachi11} by a factor $S_0$.
\end{references}
\end{document}